*Vlachokyriakos, Vasilis, Johnson, Ian, Anderson, Robert, Claisse, Caroline, Zhang, Viana, Briggs, Pamela (2024): Design Implications for a Social and Collaborative Understanding of online Information Assessment Practices, Challenges and Heuristics. In: Proceedings of the 22nd European Conference on Computer-Supported Cooperative Work: The International Venue on Practice-centered Computing on the Design of Cooperation Technologies – Notes, Reports of the European Society for Socially Embedded Technologies (ISSN 2510-2591), DOI: 10.48340/ecscw2024_n04*

# Design Implications for a Social and Collaborative Understanding of online Information Assessment Practices, Challenges and Heuristics

Vasilis Vlachokyriakos[1], Ian G. Johnson[1], Robert Anderson[1], Caroline Claisse[1], Viana Zhang[1], Pamela Briggs[2]
[1]Newcastle University, [2]Northumbria University
*[1]{vasilis.vlachokyriakos1, ian.johnson2, rob.anderson, caroline.claisse, n.zhang10}@ncl.ac.uk, [2]p.briggs@northumbria.ac.uk*

**Abstract.** The broader adoption of social media platforms (e.g., TikTok), combined with recent developments in Generative AI (GAI) technologies has had a transformative effect on many peoples' ability to confidently assess the veracity and meaning of information online. In this paper, building on recent related work that surfaced the social ways that young people evaluate information online, we explore the decision-making practices, challenges and heuristics involved in young adults' assessments of information online. To do so, we designed and conducted a novel digital diary study, followed by data-informed interviews with young adults. Our findings uncover the information practices of young adults including the social and emotional motivations for ignoring, avoiding, and engaging with online information and the ways this is entangled with collaborative arrangements with algorithms as agents. In our discussion we bring these findings in close dialogue with work on information sensibility and contribute rich insights into young peoples' information sensibility practices embedded within social worlds. Finally, we surface how such practices are attuned to prioritise wellbeing over convenience or other commonly associated sufficing heuristics.

# Introduction

Recent significant global events, such as pandemics, wars, and climate catastrophes, paired with developments in Large Language Models (LLMs) and Generative AI (GAI) have created a confused and uncertain information landscape (Ehrlén et al., 2023; Flintham et al., 2018), leaving people being increasingly susceptible to online misinformation and disinformation (Scherer & Pennycook, 2020) and other online harms (Mannell & Meese, 2022). Rising social media use, especially by young people (de Groot et al., 2023; Juvalta et al., 2023), is changing the information ecosystem with media organisations increasingly prioritising social media for news dissemination (Juvalta et al., 2023) and adopting their content and 'news values' to increase attention and 'reach' (Mast & Temmerman, 2021; Valaskivi, 2022). In this context, the lines between news and information are blurred, and researchers have developed an understanding of information seeking and assessment as relational (Borgatti & Cross, 2003) and information search as 'social' (Evans & Chi, 2008; Oeldorf-Hirsch et al., 2014).

In this paper, we build on such work by inquiring into how young adults evaluate socially encountered information. To do so, we designed an online diary study, which we conducted with young adults (age 18-35), followed by a set of data-informed interviews. We specifically targeted young adults because of known challenges in misinformation within this demographic (Pérez-Escoda et al., 2021; Qayyum et al., 2010), while also our project collaborator (the BBC), indicated that their audience demographic data point to this age group as being the one with the least engagement with media produced by their organisation across platforms. Extending recent work on 'information sensibility' (Hassoun et al., 2023) that emphasised the social ways information is sought and interpreted by young people, our findings expand insights into the decision-making practices, challenges and heuristics involved in making assessments of information online. Finally, we surface how such practices and heuristics are used to prioritise people's wellbeing, as opposed to simply being 'convenient' and 'good enough' information assessment practices (Connaway et al., 2011; Flintham et al., 2018; Hassoun et al., 2023; Metzger & Flanagin, 2013). Such insights contribute implications for designing social platforms and information ecosystems that account for such social and collaborative assessments of information online; more specifically in collaborative arrangements with algorithms and for preserving young people's wellbeing in engagements with information online.

# Background: Online Information Assessment and Sensibility

The rise and promulgation of misinformation and associated social, political, and personal risks (Das & Ahmed, 2021; Fernandez & Alani, 2018; Ruokolainen et al., 2023; Vaccari & Chadwick, 2020) has brought urgency to the challenge of understanding how people assess information online. Research on information evaluation has traditionally looked to support people to make 'better' evaluations through the design of technical interventions (Bauer et al., 2013; Flintham et al., 2018; Yang et al., 2019), focused on the online detection of misinformation (Monti et al., 2019; Shu et al., 2019; Yang et al., 2019) and relatedly, research has focused on user interface design to guide or warn 'users' of misleading information (Bauer et al., 2013; Sharevski et al., 2021) and evaluated the usefulness of online credibility indicators (Clayton et al., 2020; Metzger & Flanagin, 2013; Zhang et al., 2018) including the effect they have on information evaluation and assessment (Menon et al., 2020). However, this work, alongside other interventions, such as those that focus on information and digital literacies (Carmi et al., 2020; Jang & Kim, 2018; Wolff et al., 2016), while being attentive to what should be done to identify 'accurate' and detect 'false' information, pay little attention to complex social and contextual factors that motivate information engagement (Herrero-Diz et al., 2020; Talwar et al., 2019).

In contrast, other work in Human Computer Interaction (HCI) and Computer-Supported Cooperative Work (CSCW) has focused on identifying the factors that contribute to people's information encounters, while also developing a better understanding of how people evaluate such encountered information. Indicatively, work in this space (Connaway et al., 2011; Geeng et al., 2020) has surfaced 'convenience' as a central factor to such information encounters, while also describing the 'good enough' reasoning tactics people employ in information assessments online. Building on this work, more recent research that focused on young people between 13-24 years (Gen Z), has underlined the social practices that influence news consumption, positing such practices as *information sensibility* (Hassoun et al., 2023). Motivated by such insights, in this paper we argue that more research is needed to investigate how young people make socially informed decisions about information they engage with online. As such, we designed and conducted a study to explore the social and relational ways young adults assess information online, and contribute additional insights on young adults' information assessment practices where news is socially encountered (Edgerly, 2017; Koh et al., 2015; Strauß et al., 2021).

# Study Design and Methods

## Recruitment

The study described in this paper is part of a bigger collaborative project with the British Broadcasting Corporation (BBC). Their audience segmentation data identified young adults (18-35) in the Northeast of England as the group with the lowest engagement with media and content produced across their platforms. We recruited participants for the study through university channels and local collaborators, resulting in 150 people applying to take part in the study by filling out a questionnaire indicating their age, education level, and postcode. We selected 19 people out of this pool of participants by using as selection criteria, the age of the applicants and their postcode – as we wanted to ensure that we covered the demographic criteria of our collaborator. Of the 19 participants (11 women and 8 men) that were selected, only 14 (11 women, 3 men, aged between 21 and 34) took part in both phases of the study (diary and interview). We compensated participants £100 on completion of the study (£50 for taking part in the diary phase and a further £50 for taking part in the interview).

## Digital diary

Informed by similar studies in HCI and CSCW (Palen & Salzman, 2002; Saltz et al., 2021) we designed a web-based digital diary that allowed us to capture participants' encounters with information, while also prompting them to reflect on the context where they encounter information, the social processes involved in selecting and assessing information, and its effects. During a two-week diary period, we instructed participants to fill their diaries with information they encounter by sharing news articles, social media posts, blogs, and any other links as they read them. We provided an email address and a phone number that participants could use to share URLs, either by emailing them to us or by sending them through SMS. For any URL participants shared with us, the web-based platform opened a form including a set of questions that could be responded to through text entry, multiple choice, sliders, and added emojis, respectively (see Figure 1). We configured the questions and modes of response to be quick and easy to respond to with the aim of eliciting a mixture of objective (*Where were you? What did you do next?*) and subjective responses (*How much do you trust it? How did it make you feel?*) around the context and the social ways that information was encountered and assessed.

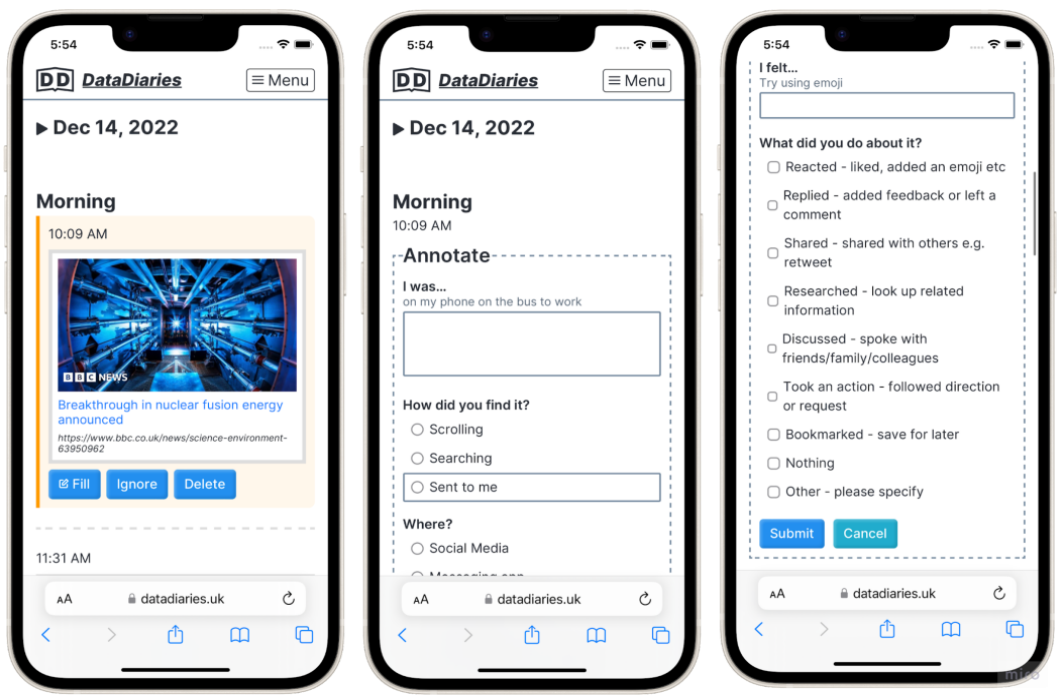

Figure 1. The online diary platform showing diary entry and annotation pages.

Data books and interviews

The data collected through the diary were not considered as primary research data, but rather as a process to sensitise participants to some of the themes of the study, encourage self-reflection in our participants, and used to create personalised Data Books to support participant interviews. The Data Books were inspired by previous design-led methods in HCI, which foster experimental engagement with personal data (Elsden, Chatting, et al., 2017; Ståhl et al., 2009). More specifically, we drew from work on designing documentary informatics (Elsden, Durrant, et al., 2017) that explores documentary uses of quantified data to support reminiscing and sense making about oneself. As such the Data Books we created, were used in conjunction with the interviews as a boundary object or 'tickets for talk' (Elsden et al., 2016).

The Data Books represented information collected for each participant during the diary phase, showing in graphic form the times of day participants shared their links with us through the platform, other factual data such as the source of stories (from the URL), information provided through the annotation task, such as whether the information came from a messaging app, social media, etc., the context (see Figure 1, centre) and what they did after (i.e., react, share, discuss, nothing – see Figure 1, right). The Data Books also contained 'inferences' about participants based on the titles and metadata as well as other details they provided through their digital diary. We used a Large Language Model (LLM) to generate such inferences and presented this with graphic representations in the Data Books (see Figure 2). For instance, the books contained visual profiles about participants based on the data they submitted in the diary phase: a speculative summary of their interests and values (i.e., 'Awareness of humanitarian issues', 'Pop culture interest') was visually represented on the fold out centre page of the Data Books (see Figure 2, right). On the last page was also featured a short poem about the participant generated by the LLM based on the other inferences and reported data from the digital diaries. As well as making the Data Books fun and engaging—we took care to create something aesthetically pleasing and presented it as a gift to the participants ahead of the interview—by using the LLM inferences and poems, we intended them to be provocative by embedding questions about privacy, algorithmic logic, and personalization.

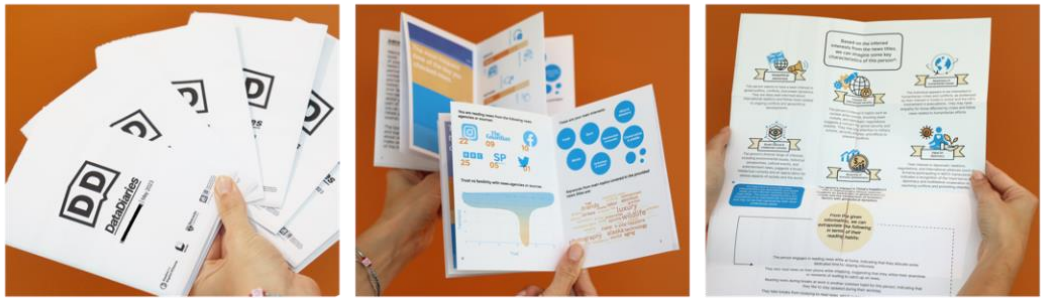

Figure 2. Photographs of examples of the Data Books given to participants and used in interviews.

Questions during the interview prompted participants to look at specific pages or graphic representations to offer explanation or provide more context on their information assessment practices. By using the AI generated inferences (e.g., generated poem, inferences about interests etc.) together with visualisations of participants reported data, we aimed to introduce a level of ambiguity in the design of the books as a resource for interpretation and speculation (Gaver et al., 2003) to promote participants' self-reflection. More specifically, the interviews (11 in person, 3 online) were structured around three themes: the first set of questions focused on the habits and behaviours drawing on what the diary process had captured; a second set of questions enquired into trust and perceptions of information veracity (also drawing on what was included in the Data Book); and a third and final group of questions focused on participants' perceptions and expectations of online news, the internet, and social media platforms.

Research data and analysis

Over a two-week period, 14 participants shared 920 links in total (range 16 - 371) from an array of sources including social media, newsletters, messaging services and online news websites and apps. The interviews lasted on average 67 minutes (range 48 – 82 mins) and were audio recorded and later transcribed. All participants signed a consent form for data collection for both diary study and interviews and the project received ethical approval by following our university's and our funding body's ethics procedures.

The first two authors of this paper conducted an inductive thematic analysis (Braun & Clarke, 2006, 2020) on the transcribed interviews. The analysis process involved a period of familiarisation with the dataset, which involved both researchers immersing themselves in the interview data, discussing, reflecting on, and making notes and creating memos about data items and the dataset. This was followed by a process of descriptive coding and the generation of initial 'candidate' themes for review and development. What followed was an iterative process of developing and refining themes before the final analytical process of writing up the analysis, as follows. All participant names are pseudonyms.

# Findings

## Information Sensibility practices

In this theme, we report on the practices that our participants referred to as ways to assess information online. More specifically we report on information validation tactics such as 'sufficing' behaviours (e.g., Simon, 1990), crowdsourcing credibility and using search engines to cross-reference news, as well as the impact of familiarity and alignment with personal worldviews in accepting the veracity of online information. However, we found that this learned behaviour needs to be re-learned and adapted in life transitions that exposed our participants to new sources and types of information.

The people in our study reported spending more time investigating and challenging what they read only if they considered it 'serious' enough, or if it did not fit with their current held worldview. For example, despite reading celebrity news for fun and entertainment, after reading unpleasant rumours about a TV personality that she liked, Hannah engaged in more rigorous research: *"I then spent the time watching more reports and videos about where people were getting their source from, because I was like, Oh, he seems like quite a nice guy."* (Hannah)

Validation tactics included 'sufficing' behaviours (e.g., Simon, 1990), and crowdsourcing (e.g., Evans et al., 2010). For example, Florence relied on the judgements of others, despite not knowing them: *"Maybe if it was something really serious, I would even look at the comments, if anyone's commented on it being a real thing."*. In addition, participants adopted a range of other 'good enough' tactics that relied on how the information 'felt' or 'seemed' – searching for other information on Google or adopting other forms of cross referencing: *"I* *think I would put in quite a general term [on a search engine]. Maybe if what I was looking for didn't come up immediately, I would say, "I'm not that bothered." Then I wouldn't search for it further."* (Lynne)

Validation processes often reflected the extent to which the information was personally relatable: *"I'd be likely to read it and sort of make my judgement from the way that it's written, I think. It would still come down to the article, how it's written, how much opinion I feel is in there."* (Hannah). Familiarity was also important. Corina for example, talks about checking for a familiar (what they consider mainstream) source to validate something seen from a less familiar source on social media: *"If I haven't seen anything in the mainstream media but see something mentioned on Twitter, that's when I think, "Oh that doesn't sound right, that doesn't match up with other things that I've read. And I guess that's kind of the way that I'm likely to check it, is to separately then go onto those other websites- or not- well, the apps, the other apps, and I see, "Oh actually, are they writing anything about it? Or is it just something new that's cropped up?"* (Corina)

People were more likely to trust information from familiar sources or that aligned with their worldview. For example, Aliza talks about how a particular news source creates suspicion based on prior experience: *"And if I look at something that isn't something that I usually go to, I trust it less. Like The Chronicle (local print and online 'newspaper'), which I do occasionally look at, but I am always, kind of, like, "Hmm. We'll see."* Julie, on the other hand shows, while beginning to reflect upon and question their own decision-making, how this familiarity heuristic also applies to the ideas or ideology: *"I don't know why I just assume it's correct. I don't know. I think it's because it's… It's probably because their politics seems to be quite similar to my own. So, because it's backing up what I believe and think, I'm like, "Oh, this must be right," which is probably a bit stupid."*

These tactics, however, must be re-learned anew during certain life transitions or when there is a cause to engage with new forms of information. For example, when becoming a new parent, Fran felt they had become less sure of using existing experiences and tactics to make decisions and judgements: *"I think, at the moment that's something that does happen to me, quite a bit, because I'm going into a new phase of my life where I'm going to be a parent, and therefore I'm more easily click-baited by stuff I like."* (Fran). Trust here is complex and reflects the trustworthiness of the news source, previous expectations around content but also confidence on one's ability to make reliable trust judgements.

## Convenience as emotional wellbeing

In this theme, we touch on the reasons why young adults sometimes engage in information that they don't trust, describing the dilemma that shapes such sceptical acceptance and that points to a willingness to accept uncertainty and a loss of control as a price for convenience, enjoyment, and emotional protection from unwelcome information.

Despite showing an awareness of the downsides with social media as a source for news, participants also admitted to being reliant on them: *"I quite like it. I think there's a lot of, "How much are they listening to? How much do they know?" But like at the end of the day it makes my experience more enjoyable."* (Hannah). Most people showed an awareness of the concerns but articulated an acceptance: *"I think as I've grown up with social media being more and more used for everything. I think it makes the news quite polarised. It's quite pessimistic, you see most of the bad things. Inciting hate and wanting people to engage in things, so making them as clickbait-y and controversial as possible. I don't like that's where social media has brought news to, but that is still the place I go to for news."* (Lynne). Part of this awareness was an acceptance that there was a limit to what could be controlled. For example, Julie described a fear of becoming accidently aware of things: *"I think my friend put it best. He said, "Everything I've ever learnt about Meghan and Harry* [i.e. members of the British Royal family] *has been against my own will," which I kind of agree with. It's like, I have never sought out information on those people, but I know more about their marriage than I do about my parents' marriage."* (Julie)

However, people were prepared to accept the negative aspects because they saw value in the information. People had clear ideas about the way they wanted to interact with information: *"An app or a website, it's just set up differently, or in the way that I think of it. Where it has these top five, top ten, it's more likely to have banners relating to what everyone wants to click on rather than what I want to look at."* (Agatha) They recognised that a personalised information 'feed' better catered to their emotional needs: *"I don't think there's a particular forum, at the moment, for me in my life, or enough emotional capacity to deal with some of the bigger topics, and to be having those conversations about bigger topics, at the moment."* (Fran). Participants described feeling shielded and protected, in particular around mental wellbeing: *"I think in a way, it's really good and healthy, because I get the kind of news that I'm interested in, and that's the news that I get to read, and I don't- if there's some kind of news that I don't like to consume, which is maybe about entertainment, or news and things I don't like to think about at all. And then I don't get those recommended, which is something I prefer a lot more."* (Meera).

## The algorithm as collaborative agent in assessments of information

Our findings point to how young adults saw algorithms that curate information on social media as unnerving, highly excitable, and "*getting the wrong end of the stick*". Nonetheless they perceived "*their algorithm*" as shielding them from information that they did not want to encounter, protecting their wellbeing, and as such, they tried to configure and 'collaborate' with them for their information needs.

Participants recognised that there were times when the algorithm would get it wrong: *"If I'm doing a lot of searches on food recipes, then I know that my algorithm is going to change to show me lots of other food recipes. That makes me quite happy […] The only time it gets annoying is when I've had enough of that thing, and I can't get rid of it from my algorithm."* (Hannah). They developed hypotheses about the ways in which the algorithm worked. For example, Lynne explained how she disengaged from a court case being followed on TikTok, noting the associated change in curated content: *"I think as I engaged with it less, I got shown it less because the algorithms are like, she's not interested in this video. I think people's opinions started to shift slightly as well, but it probably was because I engaged with it less. Now I'm thinking, "Why did I engage? What shifted that?" I don't know."* Despite, a feeling that one's algorithm could be understood, there were occasions when they over-stepped boundaries and came across as unnerving: *"There have definitely been times where I've noticed something and I'm like, "That is actually really creepy." I'm not massively into the idea that, like, for example, if a friend shared a link with me and then all of a sudden, I'm getting ads for something, I'm like, "That is a little bit creepy."* (Aliza)

Participants developed theories about how to 'game the algorithm': *"There are some articles that I'll see as clickbait that I just will not engage with, […] because that will impact my algorithm and then I'll get more of it. I'm like, "No, do not click that," because it can be on important topics. If it's an important topic I will not, I will not allow myself to be fully click-baited."* (Fran).

Some developed ways to engage with information without letting 'the algorithm' know about it: *"Sometimes, I'll read through the thread, because there are interesting arguments in the comments, but I wouldn't ever click on the news source. Because I don't want to give it the click and give it the validity of having gone into it and looked at the article."* (Julie)

One several occasions participants referred to behaviours designed to 'trick' their algorithm. Hannah, for example, talked about a slight of hand to bemuse her algorithm: *"Try and scroll past it as quickly as, or don't click on it. Yes, just try and not engage with it and hope that it picks up that I've had enough. Which it normally does."* Such ideas of obfuscating behaviours point to a concern about an ever-watching presence of one's algorithm, which must be trained, tamed, and even tricked to assert agency over the configuration of online spaces.

Ultimately, however, participants were prepared to forgive 'their' algorithm because of what they felt it offered them in terms of convenience, protection or shielding from things they did not want to encounter (i.e., a way to select information), as well as finding comfort and familiarity.

# Discussion: Expanding Information Sensibility

## Challenges: Misrepresentation, Lack of Algorithmic Control, and Harnessing Social Relations

Our participants were aware that their reliance on 'the algorithm' to socially curate information was not ideal. Most were resigned to the fact that they would be presented with unreliable or irrelevant information but could adopt information checking or validating tactics when they felt it was important. They expressed frustration with the algorithm when it seemed their feed didn't 'speak to them'. This occurred when information had the wrong tone or failed to reflect their interests. Our participants talked about being *misrepresented* if an issue came from an unfamiliar source, adopted an unusual perspective, or was politically unpalatable. However, participants recognised the dangers of a 'filtered universe' (de Groot et al., 2023) and understood *the need for a more holistic understanding of events and reality.*

Participants generally seemed to view '*their*' algorithms as over-excited, with a tendency to overreach, but nonetheless they felt ownership towards them. Our participants developed folk theories (Eslami et al., 2016; Medin & Atran, 1999; Toff & Nielsen, 2018) about how to train their algorithm. Tactics included scrolling quickly past 'irrelevant' news and engaging more meaningfully with information that reinforced their worldview. Sometimes the algorithm was seen as too effective, leading to participants feeling uncomfortable and 'creeped out'. In general, our findings point to the *need for more visibility in terms of how algorithmic curation works, and more explicit controls for algorithmic configuration.*

Regarding Hassoun et al.'s (2023) findings on the fear of making social errors, particularly in terms of sharing inaccurate or unpopular information and facing the social consequences of being wrong online, our participants showed less concern about being incorrect. Instead, their primary focus was on how news could *contribute to nurturing existing relationships*, both online and offline. For example, people talked about reading news in-depth to have meaningful conversations with family and friends, while others shared URLs with people to read and discuss together at home. In general, such insights point to very *practical ways that information systems can be designed to support more meaningful social relations.*

## Practices: Social, Collaborative, and Algorithmic Sense-Making

Young people adopt a range of practices to overcome their information challenges. For example, Hassoun et al (2023) report on how people seem to privilege lived experience, finding a trusted person as a go-to source, crowdsourcing information credibility, and practicing good enough reasoning. Our findings build on this work, showing how young people's informational practices are informed by a system of values and politics that in many cases are conflicting and are changing.

Our participants recognised that they needed to assess key information for its veracity. For example, our participants talked about doing their own research by searching for keywords of news (e.g., on Google) and checking what news sources are reporting on such news and how – a form of *surrogate thinking* (Toff & Nielsen, 2018). Young adults also talked about reading comments of news posted on social media and having discussions on instant messaging platforms and in-person as a form of *crowdsourcing credibility* (Ellison et al., 2013; Evans et al., 2010). In relation to crowdsourcing credibility, our findings also surface how practices such as reading comments were used as a way of avoiding clicking on news that might train the algorithm inappropriately. As such our data complements related work in this space by further complicating how such practices are shaped by not only informational needs but also the need to configure and 'train' the algorithm to 'behave appropriately'. In alignment with research into users' understandings of and relationship with algorithmic curation (Eslami et al., 2016; Karakayali et al., 2018), we suggest framing this phenomenon as the *algorithm as collaborator* in online information assessment. Such a framing helps us conceptualise the algorithm as an agent in information practices within complex social networks of information evaluation and sharing, while acknowledging that people must learn to collaborate well with their algorithm to better support their role in social networks as well as protect their own wellbeing. Our role as designers should be to make it easier for people to enact such collaborative encounters with curatorial algorithms, by enabling more meaningful bidirectional interaction and communication, though, for example, integrating intelligent agents in more explicitly (rather than through implicit inferences) stipulating how the algorithm learns.

Across our themes we see how these information processing practices are forms of 'sufficing' – *good enough reasoning* tactics used dependent on the context (e.g., news app or social media) and type of information that needs to be assessed. A novel dimension to such *sufficing tactics* used, is that they need to be relearned and *adjusted during life transitions* (e.g., during pregnancy when new types and sources of information are encountered). Such changes to priorities can have an impact on people's susceptibility to engaging with information that is not representative to what they need (e.g., misinformation). People found themselves 'making mistakes' in how they interacted with online information, which adjusted their algorithm in ways that they didn't want to that led to them encountering information that they would not otherwise choose to.

### Heuristics: Emotional Wellbeing through Prioritising the Familiar

Work on information sensibility identifies convenience, tone and aesthetic as the applied trust heuristics used to inform assessment practices (Hassoun et al., 2023). Similar to related work (Boczkowski et al., 2018; de Groot et al., 2023; Goyanes et al., 2023; Strandberg et al., 2019), our participants talked about news that "seems" or "feels" right, but when prompted further, they talked about *familiarity of source* and *familiarity of politics and values* as opposed to style and aesthetic. Such familiarity we also found that relates to people's awareness of how such information was algorithmically curated to them (Karakayali et al., 2018; Koenig, 2020; Swart, 2021) – i.e., information that "felt right" was information that made sense to people to appear in social feeds and news apps due to past interaction with the algorithm.

Our findings also indicate that people are both reflexive and reflective in their decisions about the information they encounter (see Kahneman, 2011), not as a matter of convenience, as for example pointed by Hassoun et al., (2023), but rather to protect their *emotional wellbeing* and feel *comforted*. We note, here, that the need for people to protect themselves by disengaging from certain types of news was a core strategy. The emotional weight (tone) of information encountered was one of the most significant factors in deciding whether to engage, but participants employed a range of strategies, sometimes resonating with those observed during the Covid-19 pandemic, when 'news avoidance' was recognised as a key wellbeing strategy (Mannell & Meese, 2022), employing a range of strategies for news avoidance, including the decision to avoid certain channels, avoid particular formats, or configure social media feeds in order to manage the kinds of news information encountered (Das & Ahmed, 2021; Patel et al., 2020). Departing from recent features of social media platforms that allow users to increase or decrease types of information presented to them (e.g., on Facebook, etc.), and in agreement with implications for design from previous work (e.g., Eiband et al., 2019; Rader & Gray, 2015; Swart, 2021; Velkova & Kaun, 2021 and others) we emphasise the need for the design of interfaces that gives more context-specific controls to users in information they encounter, and more transparency to the ways that algorithmic curatorial practice work when deciding implicitly to avoid or engage with certain types of information.

## Conclusion

In this paper we build on recent work on 'information sensibility' practices, challenges, and heuristics. We surface the challenges of: (i) *misrepresentation*, (ii) *lack of algorithmic control and transparency*, and the challenge of (iii) *staying informed to care for social relations*. Our findings validate recent work on information sensibility practices, by identifying young people's information assessment practices as highly social and collaborative, while however identifying '*my algorithm*' as an actor within such social and collaborative practices. Finally, our empirical findings point to the way that information sensibility heuristics are used as a means of protecting *emotional wellbeing*, specifically using convenience and tone as a protection against potentially harmful content. Factors such as *source of familiarity* and *value alignment* were considered as critical, as opposed to the secondary consideration of aesthetics. Such findings contribute additional empirical insights into young people's decision-making practices in assessments of information online, and design implications towards systems aiming to address misinformation when embedded in social contexts online.

# Acknowledgments

The work reported in this paper is part of the UKRI EPSRC Centre for Digital Citizens (EP/T022582/1). Special thanks to our collaborators from BBC R&D, Rhianne Jones for her continuous support, and Laura Ellis for her involvement at the initial exploratory stages. We would like to thank our participants for the information and news they shared with us, and for their enthusiasm during the interviews. Data supporting this publication are available under a CC BY 4.0 License at https://doi.org/10.25405/data.ncl.25488064.